\documentclass[nofootinbib,aps,letterpaper,preprintnumbers,superscriptaddress,showpacs]{revtex4}
\usepackage{amsmath}
\usepackage{eurosym}
\usepackage{amssymb}
\pdfoutput=1
\usepackage{graphicx}
\usepackage{color}
\usepackage{braket}
\usepackage{amsmath}
\usepackage{amssymb}
\usepackage{subfigure}
\usepackage{amsfonts}
\usepackage[left=1.5cm,right=1.5cm,top=1.5cm,bottom=1.5cm]{geometry}
\usepackage[usenames,dvipsnames,svgnames]{xcolor}  
\usepackage{hyperref}   
\definecolor{beamer@PRD}{RGB}{46,48,146}

\begin{document}
\newcommand\be{\begin{equation}}
\newcommand\ee{\end{equation}}
\newcommand\bea{\begin{eqnarray}}
\newcommand\eea{\end{eqnarray}}
\newcommand\bseq{\begin{subequations}} 
\newcommand\eseq{\end{subequations}}
\newcommand\bcas{\begin{cases}}
\newcommand\ecas{\end{cases}}
\newcommand{\p}{\partial}
\newcommand{\f}{\frac}

\title{On the viability of Planck scale cosmology with quartessence }

\author {\textbf{Mohsen Khodadi}}
\email{m.khodadi@stu.umz.ac.ir}
\affiliation{Department of Physics, Faculty of Basic Sciences,\\
University of Mazandaran, P. O. Box 47416-95447, Babolsar, Iran}

\author {\textbf{Kourosh Nozari}}
\email{knozari@umz.ac.ir (Corresponding  Author)}
\affiliation{Department of Physics, Faculty of Basic Sciences,\\
University of Mazandaran, P. O. Box 47416-95447, Babolsar, Iran}
\affiliation{Research Institute for Astronomy and
Astrophysics of Maragha (RIAAM),\\
P. O. Box 55134-441, Maragha, Iran}

\author {\textbf{Fazlollah Hajkarim}}
\email{hajkarim@th.physik.uni-bonn.de}
\affiliation{Bethe Center for Theoretical Physics and Physikalisches Institut, Universit{\"a}t Bonn, Nussallee 12, D-53115 Bonn, Germany}


\begin{abstract}
Growing evidence as the observations of the CMB (cosmic microwave background), galaxy clustering and
high-redshift supernovae address a stable dynamically universe
dominated by the dark components. In this paper, using a qualitative
theory of dynamical systems, we study the stability of a unified dark
matter-dark energy framework known as quartessence Chaplygin model (QCM) with three different
equation-of-states within ultraviolet (UV) deformed
Friedmann-Robertson-Walker (FRW) cosmologies without Big-Bang
singularity. The UV deformation is inspired by the non-commutative
(NC) Snyder spacetime approach in which by keeping the transformation groups and rotational symmetry there
is a dimensionless, Planck scale characteristic parameter $\mu_0$
with dual implications dependent on its sign that addresses the required invariant
cutoffs for length and momentum in nature, in a separate manner. Our
stability analysis is done in the $(H,\rho)$ phase space at
a finite domain concerning the hyperbolic critical points. According to our analysis, due to
constraints imposed on the signs of $\mu_0$ from the phenomenological
parameters involved in quartessence models $(\Omega_m^*, c_s^2, \rho_*)$,
for an expanding and accelerating late universe, all three QCMs
can be stable in the vicinity of the critical points. The requirement of
stability for these quartessence models in
case of admission of a minimum invariant length, can yield
a flat as well as non-flat expanding and accelerating universe in which Big-Bang singularity
is absent. This feedback also phenomenologically credits to braneworld-like
framework versus loop quantum cosmology-like one as two possible scenarios
which can be NC Snyder spacetime generators (correspond to $\mu_0<0$ and $\mu_0>0$, respectively).
As a result, our analysis show that between
quartessence models with Chaplygin gas equation-of-states and accelerating
FRW backgrounds
occupied by a minimum invariant length, there is a possibility of viability.
 \end{abstract}
\maketitle

\section{Introduction}

For decades particle physicists and cosmologists have focused on
beyond standard model physics and modified gravity theories to achieve
a clear understanding
of the character of two mysteries and challenges of standard cosmology
in our age i.e \emph{``dark matter''} (DM henceforth) and
\emph{``dark energy''} (DE henceforth). Despite the fact that none of them
has any explicit evidence in laboratory physics, these two theories can separately
 provide a consistent explanation of surprising results indicated by the
current astronomical observations \cite{SNe1}-\cite{WMAP2}. Specifically, DM initially
was suggested to explain the rotation curves of
galaxies and cluster dynamics which it was not justifiable by standard baryonic
matter. Later, the proposal of DM extended to cosmology concerning on the
issue of structure formation at large scales. Also, to illustrate the accelerated
expansion of our universe, the dominated existence of an unknown component called
DE is essential. In continuing this path, unlike the original assumption that
these two theories are different from each other, an interesting idea proposed that
DM and DE can be two manifestations of a single physical entity. From the perspective
of unification, it would be interesting to verify the possibility of a single unknown component
(or field) rather than two ones which can explain the role of both. Such a unified
framework of DM-DE (with a density ratio of approximately $0.3-0.7$), in literatures
was coined to name \emph{``quartessence"}, see \cite{q1, q2, q3, q4, q5, q6, q7, q8}
for instance. The most interesting quartessence models studied so far are
based on the Chaplygin gas
model as well as its upgraded versions as an exotic background fluid with equation-of-state
different from standard perfect fluid \cite{q9, q10, q11}. There are also other relevant
equation-of-states which some authors \cite{Reis:2004, Reis:2005} have offered
them as ansatzes which in asymptotic limit cases show the same behaviors for
the background fluid.
Chaplygin gas models are constrained by
 cosmic microwave background (CMB) and other astrophysical experiments
  \cite{Bento:2003dj,Bean:2003ae,Marttens:2017njo,Aurich:2017lck}.
However, there are some valid regions of parameter space
which motivates us to consider them as consistent
models with the current experimental data.
These regions for an equation of state like $p\propto -1/\rho^n$
with some spacial values of $n$ are not excluded, so that it can still behave
like a matter component at early eras and a cosmological constant
at late times \cite{Zhai:2017vvt,Marttens:2017njo,Aurich:2017lck}.
Additionally, these models as candidates for
dark energy are free of fine-tuning problem that appears in
the standard cosmology and quintessence models \cite{Bento:2003dj}.
Also, they can elevate the cosmic coincidence problem that appears in the relatively
constant ratio of relic density of cosmological constant and
matter content of the universe \cite{Capozziello:2013wha}.
These models can explain the formation of large structures in the universe and
the halo of DM in galaxies \cite{Beca:2003an,deBeer:2017jla}.
Even if the origin of DM and DE are different, such models can be used as a simplified model to study
all matter content of the universe as a single fluid illustrating the cosmic evolution \cite{Davari:2018wnx}.
\\

However, in theoretical physics community this paradigm is ruling
to get a complete and coherent view from the early moments until late
universe, a quantum description of early moments of
cosmology in the absence of micro-level singularity, is required.
This is despite the fact that quartessence models as $\Lambda$CDM
(cosmological constant and cold dark matter) are
based on standard cosmology which suffers from initial conditions
issue in particular past singularity. Clearly, an initial singular
state with infinite values of physical quantities, such as temperature
or energy density, should be excluded from any cosmological model.
The prevailing belief is that quantum gravity (QG) settings as
a framework which explores the universe at the micro-level spacetime
(Planck scale), are natural solutions for solving this issue.
So far, many intellectual efforts done by QG community has
led to the view that at micro-level, spacetime continuum breaks down
into a discrete one. So that even disjointed foam is very hard to
peace with the GR principles due to the fact that it should endure a later
transition to a spacetime continuum, \cite{Burgess:2003}. Despite the fact
that most known proposals related to QG such as loop quantum gravity
\cite{LQG1, LQG2}, string theory \cite{ST1, ST2}, deformed special relativity
\cite{DSR1,DSR2}, are currently at a development stage, they predict qualitatively
a different spacetime beyond some characteristic scales such as Planck length (energy)
and momentum. Therefore, in these models, Planck scale through
separation of full quantum spacetime from classical one, acts as a
natural border line or cutoffs which leads to the appearance of
some corrections in the high energy physics. Indeed, the above
mentioned invariant scales induce some extensions of the standard
uncertainty relation (Heisenberg uncertainty principle) so called ``generalized uncertainty
principle" (GUP) \cite{GUP1, GUP2} which governs the motion of particles
in micro-level spacetime. Also, need for existence
of GUP proposals at some concrete scales of distances and
energies is highly confirmed via gedanken experiments \cite{Fabio1999}.
However, as a more advanced alternative to GUP(s), there is a non-commutative
(NC) spacetime \cite{NC1, NC2} idea arising from the
results of string theory in which moreover discarding the point
like concept of the structure of spacetime can also be viewed as NC
by changing the nature of the spacetime coordinates. Given that
for each of the existing QG proposals, there is a relevant version
of GUP. So it is important to mention that some GUPs, particulary
generalized algebras designed by Kempf et al. \cite{Kempf1, Kempf2} via
offering the possibility of space quantization, are compatible
with NC spaces. One of the outstanding achievements of NC spacetime
idea which is required to get a consistent framework of QG is that
it leads to the removal of the paradox appeared due to the creation
of a black hole for an event that is sufficiently localized in spactime,
\cite{Dop:1994} see also discussions displayed in
\cite{Douglas:2001, Nicolini:2009, Anna:2011, Anna:2013, Mohsen:2016}.
Also, by attaching the NC space idea to standard quantum field theory
some positive feedbacks have been extracted. For instance, the singular
behavior of the Einstein equations in very micro-level distances has cured
within the NC space based on quantum field theory. \\

With this preface, in the present paper, through employing the methods
of qualitative theory of dynamical systems \cite{Dynamic1, Dynamic2} we
want to study the stability of a cosmology with GUP relevant to Snyder
NC deformed Heisenberg algebra \cite{Sny}. Moreover, we assume the background fluid
is supported by quartessence Chaplygin models (QCMs). We have selected the
Snyder NC space approach since it can be connected to some
``deformed special relativity'' models released in \cite{Gilkman1:2002, Gilkman2:2003}
as well as it has some incentives from loop quantum gravity \cite{Livine:2004}.
Another advantage of the underlying QG proposal for extending it into
cosmology setup is that it respects rotational symmetry,
unlike some of its other counterparts. Also within extended
framework at hand the Big-Bang singularity can be absent due to
bouncing mechanism induced by quantum correction terms in Friedmann
dynamic equations \cite{Batti}. The aforementioned positive feature
concerning the resolving of initial condition problem from one side
and valuable phenomenology functionality of the QCMs at large scales
along with this fact that they are stable into standard cosmology
\cite{Szy,Velas}, from other side, motivates us to explore the response
of this question: \emph{``Whether the quartessence Chaplygin cosmologies (QCCs) are still
stable in a free initial singularity cosmological framework suggested
by Snyder NC space approach to QG?"}. The result would be desirable in case of
``yes". Since it means that the QCCs are also able to justify
the current observations of a universe which has not been raised from a
Big-Bang singularity. Of course in light of study done in \cite{khosravi I2008, khosravi II2008}
we know that a fundamental cutoff as minimal length can play the role of dark energy
(especially cosmological constant) at late time cosmology. Also recently, people shown
that in the context of loop quantum cosmology, by taking an infrared natural cutoff within
standard FRW cosmology,  there is a possibility to explain the current
acceleration of our universe too\footnote{It should be noted that
some people try to remove the need for a mysterious matter and energy in nature through
modified gravitational theories. However, the recent measurement of the speed of gravity
with the gravitational wave ruled out many modified gravity theories as alternative
explanation to dark energy \cite{PLB2017}.}\cite{PLB2016}. However, in what follows by
admitting the existence of dark components in universe as the most common and challenging
paradigm in modern cosmology which has been able to provide successful justifications of the the
stability of galaxies and also observational data, then it turns out that the background fluid
quartessence models can be consistent within a QG extended cosmological framework which is
free of micro-level singularity. This consistency can have a dual function.
First, it can be interpreted as a
step towards providing a coherent theoretical picture from the beginning to the present day.
Second, it will be seen that the conditions
of $H>0$ and $c_s^2>0$ which refer to expanding and accelerating universe
lead us to admission and subsequently rejection of some theoretically
possibilities for the Snyder dimensionless characteristic parameter $\mu_0$
via its connection with phenomenological parameters involved in underlying
QCMs. More exactly, the present paper qualitatively suggests
the possibility of control of the behavior of Planck scale characteristic parameter
via the current astronomical signatures which indicates a down to up phenomenological
view.

\section{Deformed FRW cosmologies from Snyder-deformed Heisenberg algebras }

Until the end of this section, we will derive Snyder deformed dynamics
equations of the FRW cosmologies. More exactly, we will regard the corrections
appeared from the NC Snyder background within the standard HUP, on the classical
trajectory of the universe. Therefore, let us first start with a quick overview
of the Snyder-deformed Heisenberg algebras by taking into account some details
required.

\subsection{Snyder-deformed Heisenberg algebras}

By concerning on an $n$-dimensional NC deformed Euclidean space,
the structure of the commutator between the coordinates can no
longer be trivial rather it is deformed as follows
\be\label{e0}
[\tilde q_i,\tilde q_j]=\mu M_{ij}~~~~~\{i,j,...\}\in\{1,...,n\}~,
\ee so that $\tilde q_i$'s denote the NC coordinates.
Here, $\mu$ points to the NC Snyder deformation (or characteristic)
parameter which its dimension and value is a squared length and a
real number, respectively. By demanding two conjectures, we will
deal with the (Euclidean) Snyder space \cite{Sny}. First, the rotation
generators $M_{ij}=-M_{ji}=i(q_ip_j-q_jp_i)$ fulfill the usual $SO(n)$
algebra as well as the translation group remains undeformed (i.e.
$[p_i,p_j]=0$). Secondly, under $SO(n)$ rotations the NC coordinates
transform as vectors which results in keeping the rotational symmetry.
In the language of algebra the second assumption translates as follows
\bea\label{e1}
[M_{ij},\tilde q_k]&=&\tilde q_i\delta_{jk}-\tilde q_j\delta_{ik}, \\\nonumber
[M_{ij},p_k]&=&p_i\delta_{jk}-p_j\delta_{ik}~.
\eea
However, it is very important to stress that there are countless number of
commutator relations between $\tilde q_i$ and $p_j$ which all of them are unanimously
 adapted to the relations (\ref{e1}). By rescaling of the NC coordinates
$\tilde q_i$ in terms of variables used in common phase space i.e. ($q_i,p_j$),
one gets a deeper understanding of the subject. By referring to works released in
\cite{Mel, Me2, Me3}, we offer the most general $SO(n)$ covariant realization for
$\tilde q_i$ as follows
\be\label{e3}
\tilde q_i=q_i\varphi_1(\mu p^2)+\mu(q_jp_j)p_i\varphi_2(\mu p^2)~,
\ee so that $\varphi_1$ and $\varphi_2$ represent two finite functions and
also the convention $a_ib_i=\sum_i a_ib_i$ is compatible. It is trivial that
to restore the standard Heisenberg algebra (i.e. $\mu=0$)
the boundary condition
$\varphi_1(0)=1$, should be administered. Note here the two functions $\varphi_1$ and
$\varphi_2$ are not unique, at all. Indeed, for any given function $\varphi_1$ which
satisfies the boundary condition $\varphi_1(0)=1$, there is a relevant function as
$\varphi_2$ which is characterized via the relation $\varphi_2=(1+2\dot\varphi_1
\varphi_1)/(\varphi_1-2\mu p^2\dot\varphi_1)$ so that $\dot\varphi_1=d\varphi_1/d
(\alpha p^2)$, see Ref. \cite{BatMel08}. So the aforementioned realization of
$\tilde q_i$ (i.e (\ref{e3})) addresses the following commutator relation between
$\tilde q_i$ and $p_j$
\be\label{xpcom}
[\tilde q_i,p_j]=i\left(\delta_{ij}\varphi_1+\mu p_ip_j\varphi_2\right)~,
\ee where results in such a GUP model for the Snyder NC space at hand
\be\label{unrel}
\Delta\tilde q_i\Delta p_j\geq\f12|\delta_{ij}\langle\varphi_1\rangle+\mu\langle
p_ip_j\varphi_2\rangle|~.
\ee
The above commutator relation along with inequality, obviously imposes that the
standard framework can be recoverable by setting $\mu\rightarrow0$. Interestingly, unlike
three dimensional systems which we deal with countless realizations of the algebra
and subsequently different GUPs (\ref{unrel}), for one-dimensional systems, there
is no such an issue. By concerning on the one-dimensional systems
the symmetry group
is trivial i.e $SO(1)=\text{Id}$ and the most general realization can be written
as $\tilde q=q\varphi(\mu p^2)=q\sqrt{1-\mu p^2}$ which makes the commutation relation
(\ref{xpcom}) and inequality (\ref{unrel}) to be re-expressed as
\be\label{xp}
[\tilde q,p]=i\sqrt{1-\mu p^2}~,
\ee
and
\be\label{uncrel}
\Delta\tilde q\Delta p\geq \f 1 2|\langle\sqrt{1-\mu p^2}\rangle|~,
\ee respectively. It should be noted that to fix the sign of the
Snyder deformation parameter $\mu$, there is a freedom. Precisely,
in case of $\mu>0$ a natural cut-off as $|p|<\sqrt{1/\mu}$ appears
on the momentum while $\mu<0$ derives an observable
minimal length for $\tilde q$ from the uncertainty relation (\ref{uncrel}).
As a noticable
result, in case of negative sign for $\mu$ at the first order, one gets the
inequality $\Delta q\gtrsim(1/\Delta p+l_s^2\Delta p)$ which is the same
thing predicted by string theory (here $l_s$ refers to string length which
can be detected with $\sqrt{-\mu/2}$), \cite{St1, St2}. In conclusion,
the Snyder-deformed commutator relation (\ref{xp}) addresses the existence
of a fundamental cut-off as maximum momentum or minimal length if $\mu>0$
or $\mu<0$, respectively.

\subsection{Snyder-deformed dynamical equations }

By turning to above review of the Snyder NC algebra, we are going
to extract the relevant deformed dynamics of the FRW cosmological models.
Indeed, we want to derive classical dynamical equations ruling the universe
which is affected by one of the possible initial corrections such as Snyder NC geometry
(the corrections come from the algebra (\ref{xp})). The classical Poisson bracket
representation of the quantum-mechanical commutator (\ref{xp}) is
\be\label{pm}
\{\tilde q,p\}=\sqrt{1-\mu p^2}.
\ee
According to the above classical representation for any two-dimensional phase space
function the Snyder deformed Poisson bracket can be re-expressed as\footnote{Deformed Poisson
bracket should meet some natural conditions which the quantum mechanical commutator
possesses as anti-symmetricity, bilinearity and
satisfies the Jacobi identity as well as the
Leibniz rules.}
\be
\{F,G\}=\left(\f{\p F}{\p\tilde q}\f{\p G}{\p p}-\f{\p F}{\p p}\f{\p G}{\p\tilde q}\right)
\sqrt{1-\mu p^2}~.
\ee
It is thus expected that the time evolution of the coordinate and momentum with respect to
Hamiltonian $\mathcal H(\tilde q,p)$ can be deformed as
\be
\dot{\tilde q}=\{\tilde q,\mathcal H\}=\f{\p\mathcal H}{\p p}\sqrt{1-\mu p^2}, \qquad \dot
p=\{p,\mathcal H\}=-\f{\p\mathcal H}{\p\tilde q}\sqrt{1-\mu p^2}~.
\ee
Now, we expand the underlying framework to the cosmological context in particular
FRW cosmological models with the following spatially isotropic metric
\be\label{frw}
ds^2=-N^2dt^2+a^2\left(\f{dr^2}{1-kr^2}+r^2d\theta^2+r^2\sin^2\theta d\phi^2\right)~,
\ee where the lapse function $N=N(t)$ and scale factor $a=a(t)$.
Also, its matter section obeys fluid energy conservation equation
\be\label{e-fluid}
\dot{\rho}+3H(\rho+p)=0\;,
\ee with a generic matter energy density $\rho$ and pressure $p$. In line element
(\ref{frw}), depending on the symmetry group, the curvature constant $k$ can be fixed to
$0$, $+1$ and $-1$ by pointing to the spatially flat, closed and open universe,
respectively. In order to compute the dynamic of the underlying FRW models the following scalar constraint should be satisfied
\be\label{scacon}
\mathcal H=-\f{p_a^2}{12a}-3ak+a^3\rho=0~,~~~~~~8\pi G\equiv 1~,
\ee
where its extended representation takes the following form
\be\label{extham}
\mathcal H_E=\f{N}{12}\f{p_a^2}a+3Nak-Na^3\rho+\lambda\pi~.
\ee
Here, $\lambda$ and $\pi$ denote a Lagrange multiplier and the momenta conjugate
attributable to $N$. By turning to the Poisson bracket (\ref{pm}), we can assume
that the commutator relation between the isotropic scale factor $a$ and relevant
conjugate momentum $p_a$ in the underlying Snyder-deformed minisuperspace obeys the following
from
\be\label{ap}
\{a,p_a\}=\sqrt{1-\mu p_a^2}\,,
\ee where if shutdowns the Snyder NC space deformation (i.e $\mu=0$), it comes back
to standard form $\{a,p_a\}=1$, as expected from GR based mini superspace. Now by having the
extended Hamiltonian $\mathcal H_E$ and Poisson bracket (\ref{ap}), one can obtain relevant
deformed dynamics equations in two-dimensional phase space $(a,p_a)$, as follows
\be\label{eqapgup}
\dot a=\{a,\mathcal H_E\}=\f{Np_a}{6a}\sqrt{1-\mu p_a^2}, \qquad \dot p_a=\{p_a,\mathcal H_E\}=N\left(\f{p_a^2}{12a^2}-3k+3a^2\rho+a^3\f{d\rho}{da}\right)\sqrt{1-\mu p_a^2}.
\ee
Eventually, by solving the constraint (\ref{scacon}) with respect to $p_a$ as well as
considering the first case in equation (\ref{eqapgup}) and also fixing $N=1$, the first
Friedmann equation modified by leading order Snyder NC space correction, reveals as
\be\label{e-a1}
H^2=\frac{\rho}{3}-\f{k}{a^2}-4\mu\rho^2 a^4+24\mu ka^2\rho-36\mu k^2~~.
\ee
Subsequently by taking time derivation of the expansion rate equation
(\ref{e-a1}), we arrive at
\be\label{e-a2}
\dot{H}=-\frac{\rho+p}{2}+\frac{k}{a^2}-8\mu a^4\rho^2+12\mu a^4\rho(\rho+p)
+24\mu ka^2\rho-36\mu k a^2(\rho+p)~,
\ee as second order deformed Friedmann equation.
In above equations, the correction terms arisen from the Snyder NC geometry,
are addressed with $\mu$ parameter which is connected with Planck length
$\ell_{pl}$ via $\mu\equiv\mu_0\frac {\ell_{pl}^{2}}{\hbar^{2}}$. To the
end of this paper, to facilitate our calculations the natural
unit is adopted i.e. $ \ell_{pl}$ ,$\hbar$ and $c$ are fixed to unity (as before $8\pi G\equiv1$).
Therefore, in the following we will work with dimensionless parameter $\mu_0$.
In the QG literatures it is thought that the value of this
parameter as well as other counterparts suggested by other GUP models,
must be constant of order unity.
Notably, by attaching the QG effects arisen from some common semi-classical
approaches within different branches of physics (both theoretically and
experimentally) so far for relevant dimensionless QG parameters released
some explicit upper bounds (e.g. can be mentioned to works as
\cite{constraint0}-\cite{constraint5}).
As mentioned before, the above deformed dynamical equations explicitly show us
that vanishing the $\mu$-terms leades to the restoration of Friedmann equations
in their standard form. At the end, by plugging the equation (\ref{e-a1}) to
(\ref{e-a2}), we obtain
\be\label{e-a3}
\dot{H}=-H^2-\frac{\rho+3p}{6}+12\mu_0\rho p a^4+12\mu_0 k a^2(\rho-3p)-36\mu_0 k^2~,
\ee where along with the continuity equation (\ref{e-fluid}) are two out of three
equations which forms a closed system for doing the Jacobian stability analysis of
three different versions of Chaplygin quartessence models, within the quantum cosmological
framework.

\section{ Stability of Quartessence Models in Dynamical
Systems with Natural UV Cutoff }

\subsection{Analysis Procedure}

We begin the discussion of this section with a succinct and useful preview
of our analysis method. Overall, there are two paths to provide a dynamical
analysis of a differential equation as $\dot{y}=f(y)$: first, finding the
relevant straight solutions. Second, reducing the analysis into a phase plane
for all defined initial conditions. The latter is the basis of \emph{
``qualitative dynamic analysis''} in which all possible solutions are
considered rather than analyzing an individual solution. More precisely,
in this way one reduces dynamics into a two-dimensional (2D) phase space
in which singular solutions $\dot{y}= 0 $ and also nonsingular ones are
displayed via critical points (CP) and phase curves, respectively. Using the
phase diagrams in a phase plane (2D space) we can clearly investigate
  some important issues such as \emph{``dynamical stability''}. Generally,
we are able to reduce every conventional cosmological dynamics to the 2D
phase plane with an autonomous system of equations similar to $\dot{x} = Q_1(x, y),
\dot{y} = Q_2(x, y)$, in which dot represents the differentiation with respect
to cosmic time. Through linearization of the Jacobian matrix at a given CP
and extracting relevant eigenvalues \footnote{Note that the mentioned eigenvalues
are invariant creatures attributed to critical points since by changing the coordinates
$x, y$ they remain unchanged \cite{Szy}.} $\lambda_{1,2}$, will be provided
the possibility of categorizing the non-degenerated (or hyperbolic) CPs  $(x_c,y_c)$.
As a reminder, in case of the real part of both eigenvalues $\lambda_{1,2}$ be
nonvanishing at $(x_c,y_c)$, the relevant CP is non-degenerated.
With a good approximation the dynamical behaviour of the above mentioned
autonomous system in the vicinity of the CP $(x_c, y_c)$ is qualitatively
traceable via the behaviour of its linear part
\begin{eqnarray}\label{e2-a0}
\left(
 \begin{array}{cc}

 \dot{x} \\
 \dot{y} \\
 \end{array}
  \right)
 =\textbf{\emph{M}}_{2\times2}|_{(x_c, y_c)}~.~\left(
 \begin{array}{cc}
 x-x_c \\
 y-y_c \\
 \end{array}
  \right)~~,~~~~~~~~\textbf{\emph{M}}=\left(
 \begin{array}{cc}
 \acute{Q_{1,x}} & \acute{Q_{1,y}}\\
 \acute{Q_{2,x}} & \acute{Q_{2,y}}\\
  \end{array}
  \right)\,,
\end{eqnarray}
where after integration, the above system gives the following solution
\begin{eqnarray}\label{e2-a1}
\begin{array}{ll}
x-x_c=Re\Big(A_1 \exp(\lambda_1t)+A_2\exp(\lambda_2t)\Big)~, \\\\
y-y_c=Re\Big(A_1 k_1 \exp(\lambda_1t)+A_2 k_2\exp(\lambda_2t)\Big)~,
\end{array}
\end{eqnarray} with $k_1=\frac{\lambda_1}{\acute{Q_{1,y}}(x_c, y_c)}
-\frac{\acute{Q_{1,x}}(x_c, y_c)}{\acute{Q_{1,y}}(x_c, y_c)}$
and $k_2=\frac{\lambda_2}{\acute{Q_{1,y}}(x_c, y_c)}
-\frac{\acute{Q_{1,x}}(x_c, y_c)}{\acute{Q_{1,y}}(x_c, y_c)}$.
Here, the prime sign refers to the derivative in terms of variables
$x$ and $y$.
 \\

By specifying the sign of the trace and the discriminant within the
2D flows dynamical systems then the possibility of a solution for
stability analysis, will be available \cite{Anos:1980}.

\begin{itemize}
  \item In case of $Det\, \textbf{M} > 0$ and the discriminant: $\textbf{D} =
  (Tr\, \textbf{M})^2 -4\,Det\,\textbf{M} >0$,  the eigenvalues are real with the
  same sign which addresses the critical point as a node. If $Tr\, \textbf{M} > 0$,
  the critical point is an unstable node i.e. a repeller or source, while if
  $Tr\, \textbf{M} < 0$  it is a stable node i.e. an attractor or sink.

  \item In case of $Det\, \textbf{M} > 0$ and the discriminant: $\textbf{D} =
  (Tr\, \textbf{M})^2 -4\,Det\,\textbf{M} <0$, the eigenvalues are complex conjugates
  which address the critical point as a focus. If $Tr\, \textbf{M} > 0$ it is an unstable
   focus while if $Tr \,\textbf{M} <0$ it is a stable focus. Also, note that if eigenvalues
   are purely imaginary then the critical point is a stable neutral center.

  \item In case of $Det\,\textbf{M}< 0$, the eigenvalues are real with opposite signs
  which address the critical point as a saddle point.

  \item In case of $Tr\,\textbf{M}=0$ and $Det\, \textbf{M} > 0$,
   the eigenvalues of the critical points have complex values
   which address the stable neutrally center type. Otherwise, if
$Det\, \textbf{M} < 0$ then the critical point represents a saddle
point.

\end{itemize}

\subsection{Model I: Generalized Chaplygin Gas Quartessence (GCGQ)}

\begin{figure}[ht]
\includegraphics[width=6cm]{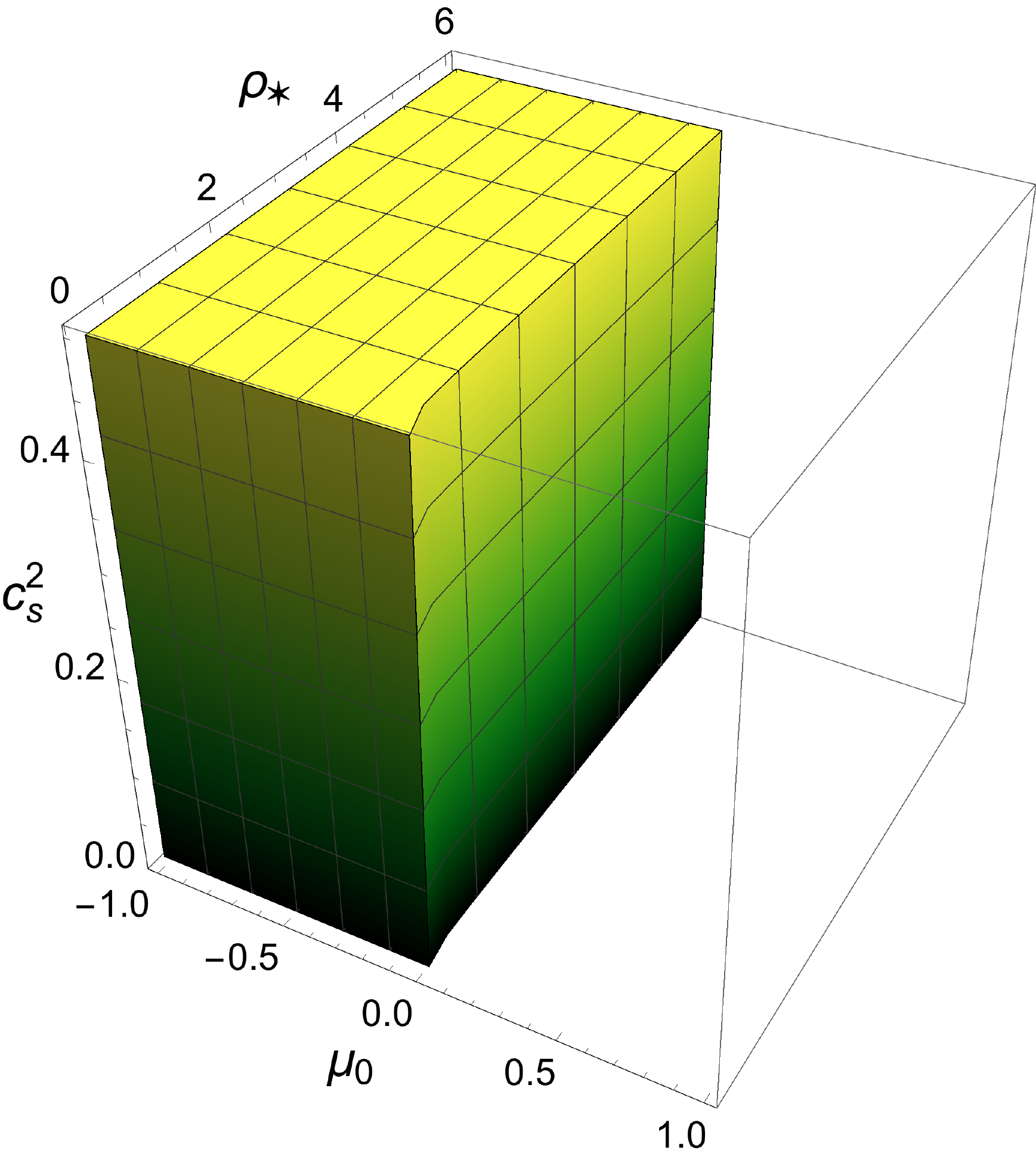}
\hspace{0.5cm}
\includegraphics[width=6cm]{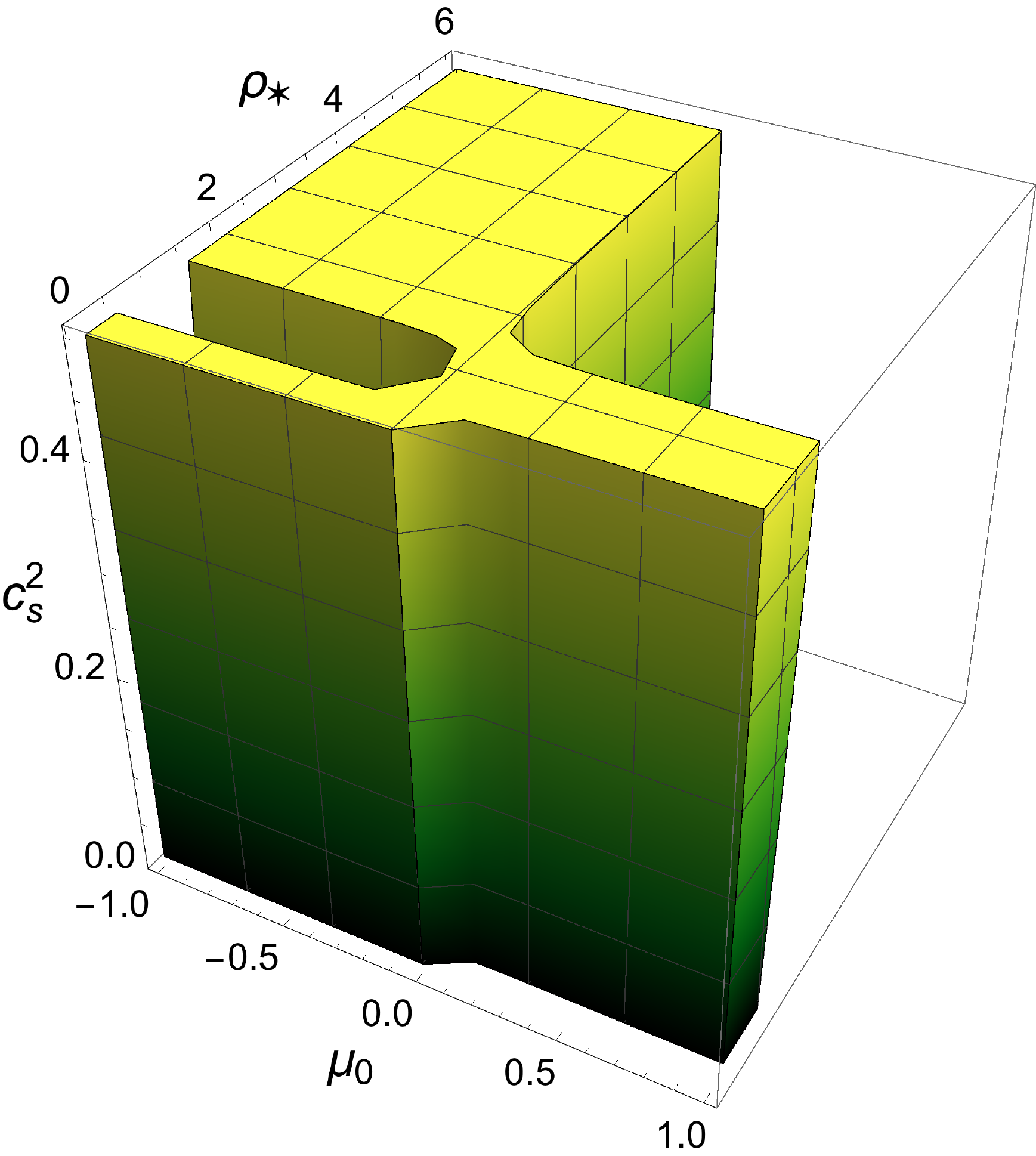}
\hspace{0.5cm}
\caption{Regions of existence $H_{cp}>0$ (Eq. (\ref{e-I10})) within $(\mu_0, \rho_*,c_s^2)$
parameter space, for flat, open (left panel) and closed (right panel) Snyder deformed
quantum cosmology with equation-of-state parameter close to $-1$ (here $\omega=-0.98$) and
any arbitrary value $\Omega_m^*\in(0.2,0.5)$.}
\label{fig*}
\end{figure}

Historically, the so called Chaplygin gas fluid model originally
studied by Chaplygin \cite{Ch:1904} in the early twentieth century
within the framework of aerodynamics via offering an exotic equation-of-state
as $p=-\frac{A}{\rho}$. However, in recent years, this model with its
upgraded versions, have been at the cosmology center of attention from
phenomenological sense so that now we see them as one of the most
popular candidates to DE-DM unified framework,
\cite{q11, Bento2, Fabris, Gorini, Bento3}. In the first
updated model of Chaplygin gas, the relevant negative pressure of
underlying background fluid is connected to energy density via
the following more general equation-of-state \footnote{This equation-of-state
and its original version (i.e. $n=1$) can be thought as a perfect fluid which
at high energy phase of universe behaves similar to a pressureless fluid while
at low energy it indicates a cosmological constant.}
\be\label{e-I1}
p=-A\rho^{-n}~,~~~~~A>0~~~~~\mbox{and}~~~~0<n\leq1 \, ,
\ee
In most literatures the above equation-of-state, describes
a ``generalized Chaplygin gas quartessence'' (GCGQ) model and
is intended as a starting point for investigations on the
cosmological implication of Chaplygin gas models. Putting above
equation-of-state into the energy conservation fluid equation
(\ref{e-fluid}), one arrives at
\be\label{e-I2}
\rho=\bigg(A+B a^{-3(n+1)}\bigg)^{\frac{1}{n+1}} \, ,
\ee for the evolution of GCGQ energy density. Here
$a(t)$ represents the cosmic scale factor which for
case of today universe can be fixed to unity.
By offering the new variables
\be\label{e-I3}
\Omega_m^*\equiv\frac{B}{A+B}~~~\mbox{and}~~~\rho_*\equiv(A+B)^{\frac{1}{n+1}} \, ,
\ee
then the equation (\ref{e-I2}) can be written as
\be\label{e-I4}
\rho(a)=\rho_*\bigg((1-\Omega_m^*)+\Omega_m^* a^{-3(n+1)}\bigg)^{\frac{1}{n+1}} \, .
\ee
Here $\rho_*$ can be interpreted as \emph{``today critical density''} of universe
since by fixing $a=1$ then $\rho(1)=\rho_*$. To provide a physical interpretation
of variable $\Omega_m^*$ it is necessary to compare the above equation with the following
$\Lambda CDM$ density energy
\be\label{e-I5}
\rho(a)=\rho_*\bigg((1-\Omega_m)a^{-3(\omega^*+1)}+\Omega_m a^{-3}\bigg)^{\frac{1}{n+1}} \, ,
\ee where $\Omega_m$ and $(1-\Omega_m)$ denote the current CDM density parameter and
dark energy density, respectively. It is clear that for spacial cases $n=0$ and
$\omega^*=-1$, these two models will meet each other which means that $\Omega_m^*$
can be interpreted as \emph{''effective matter density parameter''} in relevant Chaplygin
gas model. Now let us follow our main aim i.e. the stability analysis of GCGQ model
within the context of Snyder NC deformed quantum cosmology. By re-expressing the
Eqs (\ref{e-a3}) and (\ref{e-fluid}) as follows
 \be\label{e-I6}
\dot{H}\equiv\frac{dH}{dt}=-H^2-\frac{\rho+3p}{6}+12\mu_0\rho p a^4+12\mu_0 k
a^2(\rho-3p)-36\mu_0 k^2=Q_1(H,\rho)~,
\ee
and
\be\label{e-I7}
\dot{\rho}\equiv\frac{d\rho}{dt}=-3H(\rho+p)=Q_2(H,\rho) \, ,
\ee we define our 2D dynamical system in which the quantities $(H,\rho)$ play the role of
the phase space variables. More precisely, the evolution of the underlying system is traceable
via trajectories into $(H,\rho)$-space uniquely specified by the initial conditions
$(H_{cp},\rho_{cp})$. Therefore, in this phase space the linearization matrix $\textbf{\emph{M}}$
of the system at the around of CP $(H_{cp},\rho_{cp})$, reads off as
\begin{eqnarray}\label{e-I8}
\textbf{\emph{M}}=\left(
 \begin{array}{cc}
 \acute{Q_{1,H}} & \acute{Q_{1,\rho}}\\
 \acute{Q_{2,H}} & \acute{Q_{2,\rho}}\\
  \end{array}
  \right)_{(H_{cp},\rho_{cp})} \, ,
\end{eqnarray} where for non-static CPs $(H_{cp},\rho_{cp})$, the trace and the
determinant are obtained  as
\be\label{e-I9}
Tr\, \textbf{M}=\big(\acute{Q_{1,H}}+ \acute{Q_{2,\rho}}\big)_{(H_{cp},\rho_{cp})}~~~~~,~~~~
Det\,\textbf{M}=\big(\acute{Q_{1,H}}~.~ \acute{Q_{2,\rho}}-\acute{Q_{1,\rho}}~.~\acute{Q_{2,H}}\big)
_{(H_{cp},\rho_{cp})} \, .
\ee
Now by setting equations (\ref{e-I6}) and (\ref{e-I7}) to zero, non-static CPs
are derived as
\begin{eqnarray}\label{e-I10}
\begin{array}{ll}
H_{cp}=\bigg[\frac{\rho_*}{3}(1-\Omega_m^*)^{\frac{\omega}{\omega-c_{s}^2}}-12\mu_0\rho_*^2(1-\Omega_m^*)^{\frac{2}{n+1}}
\bigg((\frac{\omega+1}{\omega})(\frac{\Omega_m^*-1}{\Omega_m^*})\bigg)^{-\frac{4\omega}{3(\omega-c_{s}^2)}}+\\
48\mu_0 k\rho_*(1-\Omega_m^*)^{\frac{\omega}{\omega-c_{s}^2}} \bigg((\frac{\omega+1}{\omega})(\frac{\Omega_m^*-1}{\Omega_m^*})
\bigg)^{-\frac{2\omega}{3(\omega-c_{s}^2)}}-36\mu_0k^2\bigg]^\frac{1}{2} \, ,\\
\rho_{cp}=\rho_*(1-\Omega_m^*)^{\frac{\omega}{\omega-c_{s}^2}} \,~~~\mbox{with}~~~n=-\frac{c_s^2}{\omega} \,,
\end{array}
\end{eqnarray} respectively. Note that, expressions relevant to scale factor
terms in (\ref{e-I6}) obtained from mixing the equation-of-state index
$\omega\equiv\frac{p}{\rho}$ and the squared sound speed $c_{s}^{2}\equiv\frac{dp}{d\rho}$
with (\ref{e-I4}). Finally for the above non static CP, we have
\be\label{e-I11}
Tr\, \textbf{M}=-H_{cp}(3n+5)~~~,~~~Det\,\textbf{M}=6H_{cp}^2(n+1)~~~,~~~D=H_{cp}^2(9n^2+6n+1) \, ,
\ee
At first look, one may think this is exactly what has already been achieved within
standard cosmology. Therefore, Planck scale corrections induced by Snyder NC space
into FRW cosmologies does not affect standard results. However, with a closer look
one will find that the effect of UV natural cutoffs embeds into $H_{cp}$ term.
Expressions listed in (\ref{e-I11}) explicitly reflect this fact that determinant and
discriminant are always positive so that to have a stable node CP there should be
$Tr\, \textbf{M}>0$ i.e $H_{cp}>0$. In another words, in an expanding universe, the CP
(\ref{e-I10}) behaves as an asymptotically stable node. Despite that in the absence
of underlying corrections, $H_{cp}$ is trivially positive, here it should be checked carefully.
Our consideration shows that concerning late time phase of the universe i.e. fixing values
close to $-1$ for equation-of-state parameter $\omega$ and respect to standard constraints
$\Omega_m^*\in(0.2,0.4)$ into flat as well as open spatial geometry model universe
at hand, the condition of $H_{cp}>0$ holds only if $\mu_0<0$ (i.e. adoption of a minimum
invariant length in fundamental level of nature), as displayed in Fig. \ref{fig*} (left panel).
However, for case of closed universe ($k=+1$), we find that depending on the fixed values for
present critical density of universe $\rho_*$, also there is the possibility of admitting
the positive value (moreover negative values) for the dimensionless Snyder characteristic
parameter $\mu_0>0$, as revealed in Fig. \ref{fig*}
(right panel).

\begin{figure}[ht]
\includegraphics[width=6cm]{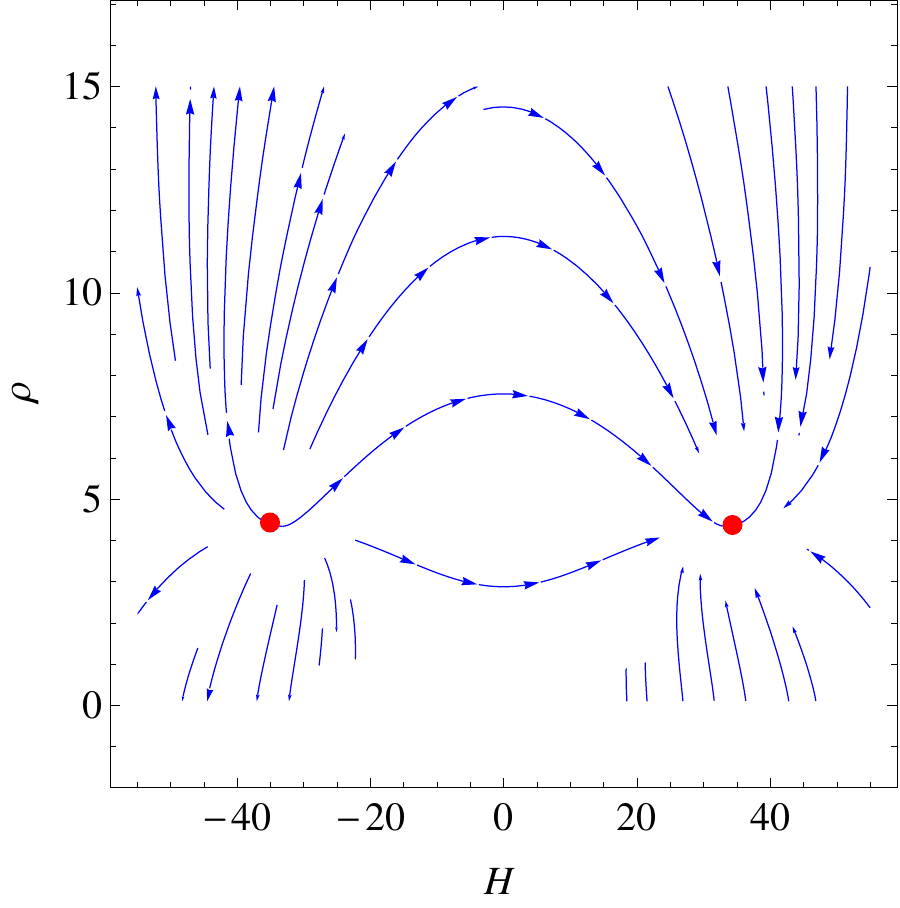}
\hspace{0.5cm}
\includegraphics[width=6cm]{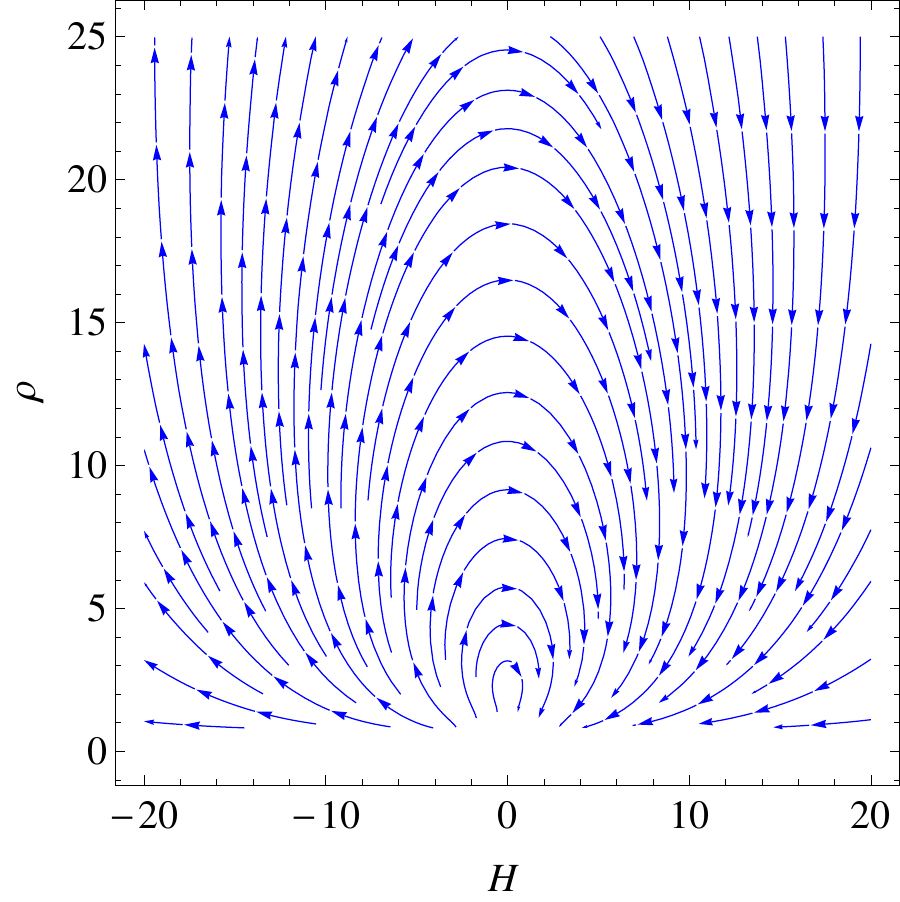}
\hspace{0.5cm}
\caption{The vector field portrait in phase space $(H,\rho)$ corresponding to Fig. \ref{fig*}.
Left panel corresponds to any three curvature modes of Snyder deformed-FRW model
with numerical values: $\mu_0=-1$,~$\rho_*=5$,~$\omega=-0.98,~\Omega_m^*\in(0.2,0.5)$,~
$c_{s}^2\in(0,0.5]$.
Right panel only corresponds to closed curvature mode with the same numerical values except $\mu_0=1$,~$0<\rho_*<2$. }
\label{pp*}
\end{figure}

Note that although in language of perfect fluid, the equation-of-state
(\ref{e-I1}) covers $-1\leq\omega\leq0$, here for all three possible modes
of spatial curvatures (i.e. $k=0,~\pm1$), the condition of $H_{cp}>0$ does not
support exactly $\omega=-1$. It is not hard to prove that the Snyder NC
space correction terms include scale factor $a$ into (\ref{e-a3}) are the
main reason of the issue so that by rejecting them this issue could be disappeared.
It is also worthy to refer that the above parameter volume addresses
interestingly the possibility of connection between two seemingly unrelated
phases of the universe. To say more exactly, the Snyder characteristic parameter $\mu_0$
deals with the earliest phase of the universe linked to the two valuable quantities
in current cosmology i.e. today critical density of universe $\rho_*$ and the
squared sound speed \footnote{As a reminder to highlight the role of this
quantity in current cosmology, note that there is a close connection
between the sign of $c_s^2$ with background dynamics of the universe. The
current accelerating phase of the universe strongly addresses a positive
sign for $c_s^2$.} $c_s^2$.
As a consequence, based on the conventional approaches to cosmology which highly
support this belief that the spatial geometry of the universe is exactly flat,
the stability of the GCGQ model within the underlying QG
extended cosmological framework will be possible only in case of admitting
a lower bound for length in nature, $\mu_0<0$. However, observational data
(primarily the CMB) tells us that the curvature constant must be close to
flat but not exactly flat. Concerning the non-flat geometries, we see from Fig.
\ref{fig*} that the behavior of $\mu_0$ for open universe is quite similar
to flat one while the sign of $\mu_0$ in closed universe is dependent on
fixed values of  $\rho_*$. Also in Fig. \ref{pp*},
it is displayed that the phase portraits in physical domain ($\rho>0$) are equivalent to terms
dictated by Fig. \ref{fig*}. As it is seen in the left panel, for each three curvature
modes of the Snyder deformed-FRW model including a minimum length, there are two de Sitter
nodes. de Sitter node in the region $H>0$ is attractor and stable, while its counterpart
in the region $H<0$ is repeller and represents an unstable CP. Concerning closed curvature mode
which includes the maximum momentum, the right panel shows circular trajectories around the static CP
$(0,\rho)$ which is affiliated to a center equilibrium CP and represents a static universe.
Note that in the left panel also one can see some static CPs associated to unstable saddle points
which are located on the trajectories moving from the unstable de Sitter node ($H<0$)
towards the stable de Sitter node ($H>0$).

\subsection{Model II: Modified Chaplygin Gas Quartessence (MCGQ)}

\begin{figure}[ht]
\includegraphics[width=6cm]{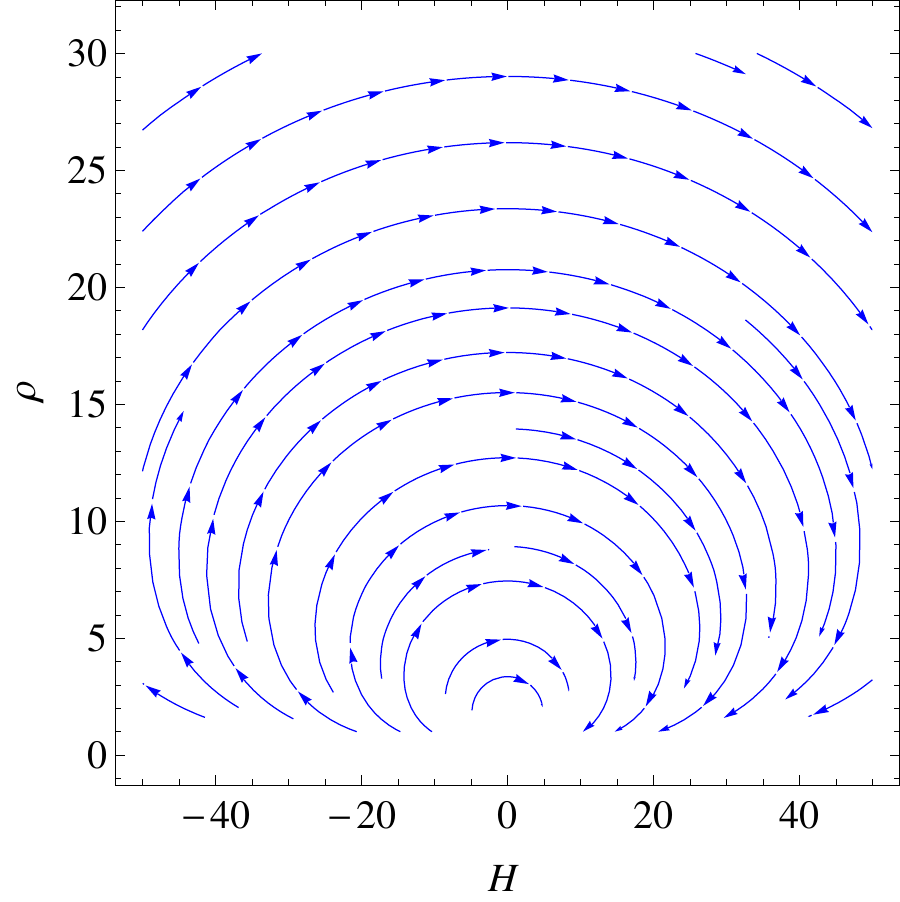}
\hspace{0.5cm}
\includegraphics[width=6cm]{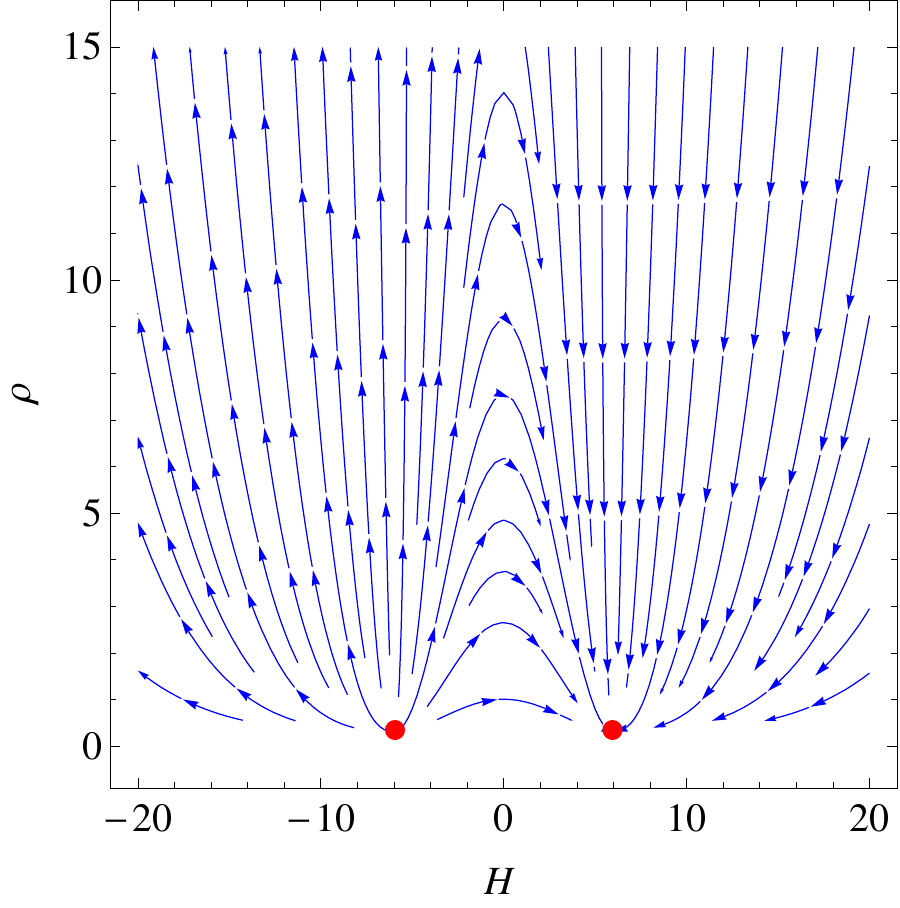}
\hspace{0.5cm}
\caption{The vector field portrait in phase space $(H,\rho)$ corresponding to Fig. \ref{fig**}.
Left panel corresponds to closed curvature mode of Snyder deformed-FRW model in the presence of
maximum momentum with numerical values: $\mu_0=1$,~$\rho_*=3,~\omega=-0.95,~\Omega_m^*\in(0.2,0.5)$,~
$c_{s}^2\in(0,0.5]$
and $C,~D>0$. Right panel is in the presence of minimum length with the same numerical values except
$\mu_0=-1,~\rho_*=6$. }
\label{pp**}
\end{figure}

Over the years, for GCGQ models several modifications have
been proposed. If one regards the modified Chaplygin gas quartessence
(MCGQ) in which pressure $p$ and energy density
 $\rho$ are connected together via the following ansatz \footnote{It
is interesting to note that, equation-of-state (\ref{e-II1}) is wider than
GCGQ model since it covers from radiation dominated era for small values
of the scale factor in the early universe to large values of the scale factor
in the late universe which cosmological constant prevails as the inducement of accelerated
expansion of our universe.} \cite{MCG1, MCG2, MCG3}
\be\label{e-II1}
p=C\rho-D\rho^{-n}~,~~~~~~C,~D>0~~~~,~~~~~0<n\leq1\, ,
\ee
then it results in
\be\label{e-II2}
\rho=\bigg(\frac{D}{C+1}+Ea^{-3(n+1)(C+1)}\bigg)^{\frac{1}{n+1}}~.
\ee
By assuming the following new variables
\be\label{e-II3}
\Omega_m^*\equiv\frac{E(C+1)}{D+E(C+1)}~~~,~~~\rho_*\equiv\bigg(\frac{D+E(C+1)}
{C+1}\bigg)^\frac{1}{n+1}\, ,
\ee
then Eq. (\ref{e-II2}) can be expressed as follows
\be\label{e-II4}
\rho=\rho_*\bigg((1-\Omega_m^*)+\Omega_m^*a^{-3(n+1)(C+1)}\bigg)^{\frac{1}{n+1}}~,
\ee
where by merging it with Eq. (\ref{e-II1}) in addition to equation-of-state index
$\omega\equiv\frac{p}{\rho}$ and the squared sound speed $c_s^2\equiv\frac{dp}{d\rho}$,
we get the following expression
\be\label{e-II5}
a=\bigg(\frac{\Omega_m^*-1}{\Omega_m^*}-\frac{D\rho_*^{-n-1}}{\Omega_m^*(\omega+1+
\frac{D\rho_*^{-n-1}}{\Omega_m^*-1})}\bigg)^{\frac{(\Omega_m^*-1)\rho_*^{n+1}}{3D(n+1)}}~,~~~~
n=\frac{c_s^2-C}{C-\omega} \, ,
\ee for the scale factor $a$  appeared in (\ref{e-a3}).
With a simple calculation one can show that in the limit $C\rightarrow0$, the above
expression reduces to its counterpart in GCGQ model. Note that, with the same argument
mentioned in details previously, here also we can interpret variables $\Omega^*$ and
$\rho_*$ as effective matter density of MCGQ model and today energy density of
the universe, respectively. Finally, in the context of quartessence model at hand, the
relevant expressions for non-static CPs, take the following form
\begin{eqnarray}\label{e-II6}
\begin{array}{ll}
H_{cp}=\bigg[\frac{\rho_*}{3}(1-\Omega_m^*)^{\frac{1}{n+1}}-12\mu_0\rho_*^2(1-\Omega_m^*)^{\frac{2}{n+1}}
\bigg(\frac{\Omega_m^*-1}{\Omega_m^*}-\frac{D\rho_*^{-n-1}}{\Omega_m^*(\omega+1+
\frac{D\rho_*^{-n-1}}{\Omega_m^*-1})}\bigg)^{\frac{4(\Omega_m^*-1)\rho_*^{n+1}}{3D(n+1)}}+\\
48\mu_0 k\rho_*(1-\Omega_m^*)^{\frac{1}{n+1}} \bigg(\frac{\Omega_m^*-1}{\Omega_m^*}-\frac{D
\rho_*^{-n-1}}{\Omega_m^*(\omega+1+\frac{D\rho_*^{-n-1}}{\Omega_m^*-1})}\bigg)^{\frac{2(\Omega_m^*-1)
\rho_*^{n+1}}{3D(n+1)}}-36\mu_0k^2\bigg]^\frac{1}{2}\, ,\\
\rho_{cp}=\rho_*(1-\Omega_m^*)^{\frac{1}{n+1}} \,.
\end{array}
\end{eqnarray}
Now, it is clear to show that, the result of (\ref{e-I11}) once again
repeats. Namely, in the presence of UV invariant cutoff raised within Snyder
NC space road to QG, the underlying MCGQ cosmological model in case of $H_{cp}>0$
(expanding universe) is stable. However, due to existence of some
correction terms, the condition $H_{cp}>0$ is not trivial rather should be checked.
As before, concerning the flat as well as open universe in late time phase,
for equation-of-state indices close to $\omega\approx-1$ (except -1) with
$\Omega_m^*\in(0.2,0.4)$, we find that independent of any arbitrary values
$C,~D>0$, the condition $H_{cp}>0$ holds only in case of $\mu_0<0$, as Fig.
\ref{fig*} (left panel). However, by taking $k=+1$ into quantum cosmological
model at hand then $(\mu_0,\rho_*,\Omega^*)$ parameter space addresses both
possibilities i.e. positive and negative signs for $\mu_0$, dependent on relevant values
for $\rho_*$, as can be seen clearly in Fig. (\ref{fig**}).
\begin{figure}[ht]
 \label{fig**}\centering
 \includegraphics[width=6cm]{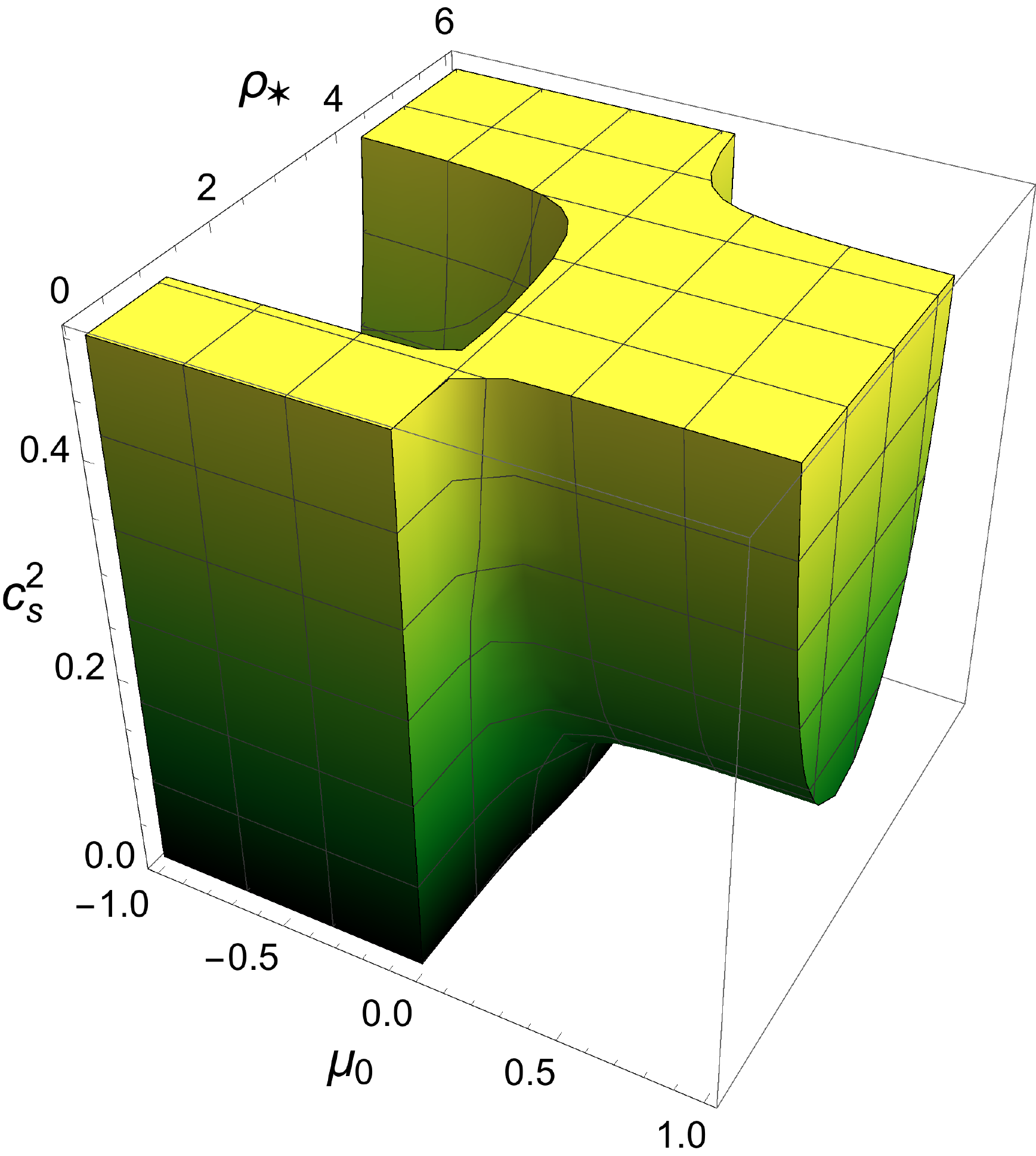}
\caption{Regions of existence $H_{cp}>0$ (Eq. (\ref{e-II6})) within $(\mu_0, \rho_*,c_s^2)$
parameter volume, for closed Snyder NC quantum cosmology with equation-of-state indexes
close to $-1$ (here $\omega=-0.95$) and any arbitrary values $n\in(0,1]$, $\Omega_m^*\in(0.2,0.5)$
and $C,~D>0$.}
\label{fig**}
\end{figure}
In similar to the former quartessence cosmology model which $H_{cp}$ has been divergent
at $\omega=-1$, here also this issue can be seen. Once again we mention that the root
of this restriction is thanks to the Snyder NC correction terms include
scale factor $a$ into (\ref{e-a3}). In Fig. (\ref{pp**})
we show the phase portraits equivalent to terms dictated by Fig. (\ref{fig**}) in the physical domain,
$\rho>0$. As before, we see that in the presence of maximum momentum there are circular
trajectories around static universe ($0,\rho$) which is affiliated to a stable center equilibrium
CP. However, in the presence of minimum length there are two de Sitter nodes which in the case of an
expanding universe, it is stable attractor while for its contracting counterpart, it is unstable
repeller. Here there is also the possibility of static universe which behaves as an unstable saddle CP.

\subsection{Model III: Modified Generalized Chaplygin Gas Quartessence (MGCGQ)}

The third proposed model for quartessence cosmology that we are interested in
here to introduce is known as modified generalized Chaplygin gas (MGCG) with
the following form of equation-of-state \cite{JCAP1, JCAP2}
\be\label{e-III1}
p=\beta\rho-(\beta+1)A\rho^{-n}\, ,
\ee where $\beta$ is an optional real constant so that in the absence of it
(i.e $\beta=0$) the GCGQ model will be recovered. It is obvious that in the hot
early universe the above equation-of-state reduces to $p=\beta\rho$ which
by fixing $\beta=1/3$ it addresses the radiation dominated epoch.  While in case of
$\beta=-1$ then $p=-\rho$, corresponding to the equation-of-state
of a cosmological constant. Here, the MGCG density evolves as
\be\label{e-III2}
\rho=\bigg((\beta+1)A+Fa^{-3(\beta+1)(n+1)}\bigg)^{\frac{1}{n+1}}\, ,
\ee where using the following new variables
\be\label{e-III3}
\Omega_m^{*}\equiv\frac{F}{A+F},~~~~~\rho_*\equiv \bigg(A+F\bigg)^{\frac{1}{n+1}} \, ,
\ee then the above MGCG density takes the following form
\be\label{e-III4}
\rho=\rho_*\bigg((1-\Omega_m^*)+\Omega_m^*a^{-3(\beta+1)(n+1)}\bigg)^{\frac{1}{n+1}} \, ,
\ee
In line with previous routes, here we arrive at the following expressions
\begin{eqnarray}\label{e-III5}
\begin{array}{ll}
H_{cp}=\bigg[\frac{\rho_*}{3}(1-\Omega_m^*)^{\frac{\beta-\omega}{c_s^2-\omega}}-12\mu_0\rho_{*}^{2}(1-\Omega_m^*)
^{\frac{2(\beta-\omega)}{c_s^2-\omega}}\bigg
((\frac{\omega+1}{\beta-\omega})(\frac{1-\Omega_m^*}{\Omega_m^*})
\bigg)^{-\frac{4(\beta-\omega)}{3(\beta+1)(c_s^2-\omega)}}+\\
48\mu_0 \rho_* k(1-\Omega_m^*)^{\frac{\beta-\omega}{c_s^2-\omega}}
\bigg((\frac{\omega+1}{\beta-\omega})(\frac{1-\Omega_m^*}{\Omega_m^*})
\bigg)^{-\frac{2(\beta-\omega)}{3(\beta+1)(c_s^2-\omega)}}-36\mu_0k^2\bigg]^\frac{1}{2} \, ,\\
\rho_{cp}=\rho_*(1-\Omega_m^*)^{\frac{\beta-\omega}{c_s^2-\omega}} \, .
\end{array}
\end{eqnarray} for the relevant non-static CPs, so that in the limit $\beta\rightarrow0$ its counterpart
in (\ref{e-I10}) can also be recovered, as expected.
\begin{figure}[ht]
\hspace{-1cm}\includegraphics[width=6cm]{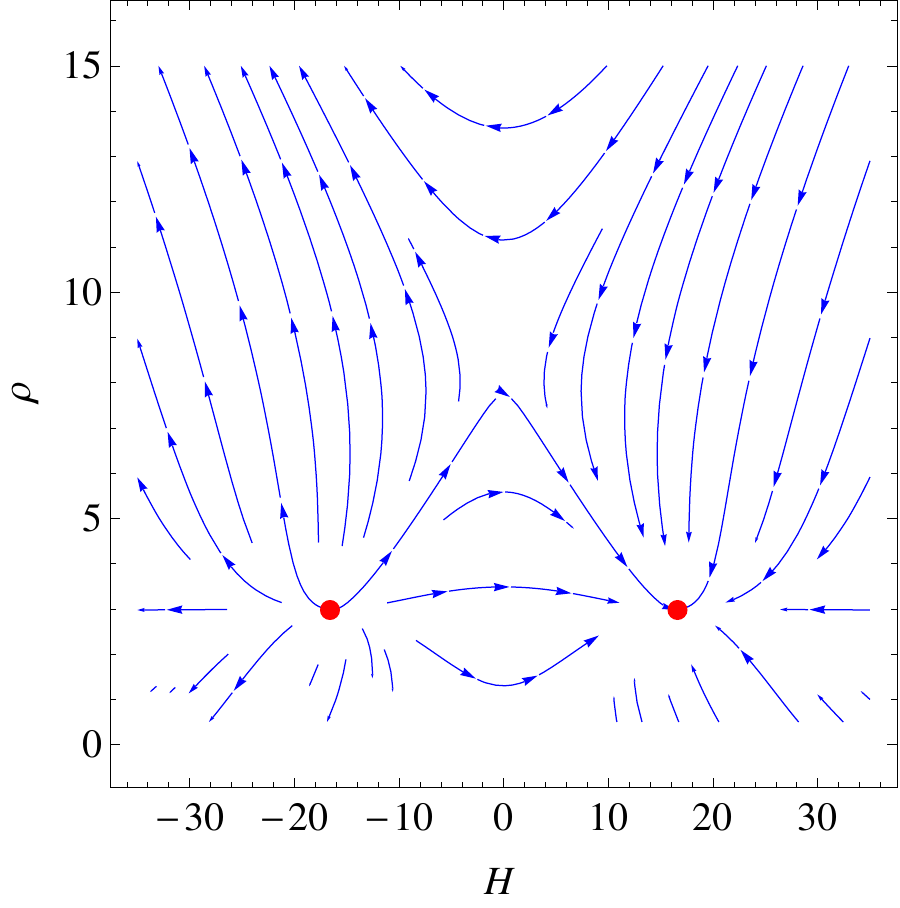}
\hspace{0.5cm}
\includegraphics[width=6cm]{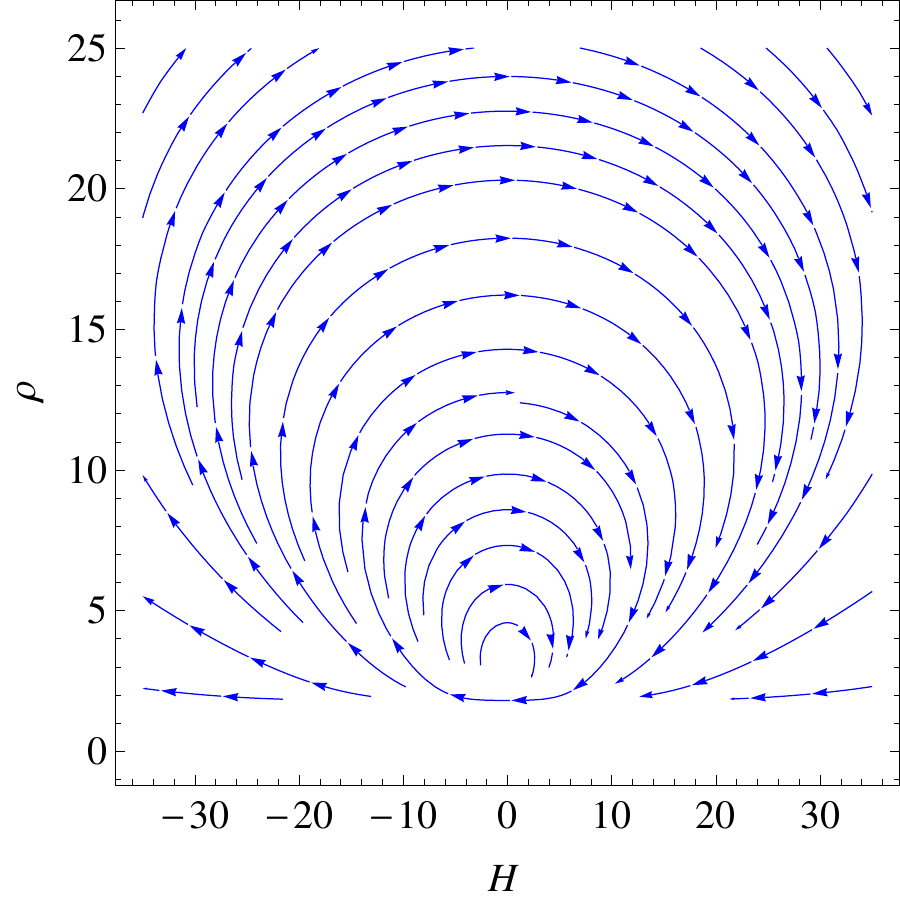}
\caption{The vector field portrait in phase space $(H,\rho)$ corresponding to MGCGQ model.
Left panel corresponds to any three curvature modes of Snyder deformed-FRW model
with numerical values: $\mu_0=-1$,~$\rho_*=5$,~
$\omega=-0.98,~\Omega_m^*\in(0.2,0.5)$,~$c_{s}^2\in(0,0.5]$ and
$\beta>-1$. Right panel only corresponds to closed curvature mode with the same numerical values except
$\mu_0=1$,~$0<\rho_*<2$.}
\label{pp***}
\end{figure}
Like the two previous models, we should follow the
validity of the condition $H_{cp}>0$ which guarantees
an expanding universe. Our analysis interestingly shows
that if the free parameter $\beta$ is in range of $\beta>-1$,
and $\Omega_m^*\in(0.2,0.4)$ then by fixing values close to $-1$ for
$\omega$, we deal with $(\mu_0, \rho_*,\Omega^*)$
parameter volumes similar to Fig. (\ref{fig*}). Namely,
for cases of flat and open spatial geometry modes
the condition of $H_{cp} > 0$  holds only in case of
adoption of a minimum invariant length in fundamental level
of nature, i.e. $\mu_0<0$. While for case of closed mode, depending on
fixed values for present critical density of universe $\rho_*$  there is
the possibility of admitting the maximum momentum and minimal length.
The remarkable thing in above results is that for all three modes $k=0,\pm1$,
the condition $H_{cp}>0$, will not be satisfied for values of $\beta\leq-1$.
In Fig. (\ref{pp***}) we draw the vector field
portraits of dynamical system relevant to MGCGQ model. Here also interpretation of the
behavior of the trajectories in the neighborhood of the CPs is similar to the two previous
models. 

\section{Concluding Discussions }

Quartessence as one of prevalent alternatives to $\Lambda$CDM,
with a phenomenologically unified dark matter-energy framework,
is based on past singular Friedmann-Robertson-Walker (FRW)
cosmology. However, in order to provide a complete picture
from the beginning of the universe to today, some ingredients
should be attached to the standard theory. In this paper, we have
focused on the stability of three quartessence models with generalized
Chaplygin gas (GCG), modified Chaplygin gas (MCG) and generalized modified
Chaplygin gas (GMCG) equation-of-state into a cosmology with generalized
uncertainty principle arisen from non-commutative (NC) Snyder space leading
to the absence of past singularity issue. The relevant dynamical equations
have been derived within a FRW minisuperspace in the presence of some invariant
UV cutoffs given by Snyder NC geometry which address a road to quantum gravity.
The UV deformed Friedman equation governing our model includes an interesting feature.
Due to freedom in the sign of the Snyder characteristic parameter $\mu$ (by setting
the natural unites ($\ell_{pl}=\hbar=c=1$) it becomes equal to its dimensionless
counterpart, i.e, $\mu_0$), then the mentioned deformed Friedman equation can be
linked to the cosmological dynamics of loop quantum gravity (LQG) by applying a
cutoff on the momentum i.e $\mu_0>0$ from one side and Randall-Sundrum braneworld
in case of a cutoff on the length i.e. $\mu_0<0$, from the other side. Using the method
of qualitative theory of dynamical systems, our stability analysis is performed within $(H,\rho)$
phase plane at a finite domain by concerning the hyperbolic critical points.
Generally speaking, for all three GCG, MCG and GMCG cases, within expanding
($H>0$) and accelerating universe ($c_s^2>0)$, the quartessence models are
stable in the neighborhood of the critical points $(H_{cp},\rho_{cp})$, in the
case of admitting one of theoretically possible signs for $\mu_0$. The
outstanding feature of our stability analysis is that it restricts freedom
to accept the expected invariant UV cutoffs via the connection between QG
free parameter $\mu_0$ and the phenomenological parameters involved in
quartessence models $(\Omega_m^*, c_s^2, \rho_*)$. In particular,
our analysis explicitly shows that the requirement of stability for
above mentioned quartessence models unanimously within a flat
accelerating universe free of Big-Bang singularity, will be
possible only in case of acceptance of a minimum invariant length in
fundamental level (i.e. $\mu_0<0$). Also, we have noticed that for all
three of the above-mentioned background fluids within the underlying
Snyder deformed cosmology with open spatial geometry, the possibility
of stability in present time only exists in case of admitting a minimum
length at the fundamental level. While for closed one, depending on the
fixed values for today critical density of universe $\rho_*$, one can accept
one of possible cases for $\mu_0$.
For any three Chaplygin gas quartessence
models, we have constructed the phase portraits in a 2D phase space
($H,\rho$) separately and discussed on the behavior of trajectories in
the neighborhood of the CPs. As a result, it is common in all three quartessence
models that in the presence of minimum length ($\mu_0<0$), there is the possibility
of a stable expanding and accelerating universe with all three possible curvature modes.
While, regarding the maximum momentum ($\mu_0>0$) within the FRW background, only shows
a stable static universe with closed spatially geometry. As a consequence, our results
are essentially independent of the free parameters of equation-of-states of Chaplygin gas models, which are constrained by experiments
\cite{Bento:2003dj,Bean:2003ae,Marttens:2017njo,Aurich:2017lck,Zhai:2017vvt}. \\

Briefly, this work contains the following important consequences.
\emph{First, the requirement of stability for three quartessence models
can yield an expanding and accelerating universe compatible with current observational
evidences in which Big-Bang singularity is absent. To be more detailed, in case of setting the flat
and open geometries for curvature constant modes within NC Snyder spacetime approach, it
will be realized the braneword-like framework along with the relevant
uncertainty relation of string theory. While for the case of closed universe depending on $\rho_*$,
also there is a chance to emerge of the LQG-like framework \footnote{In light of our results,
within the framework of flat universe which is accepted by the physics community,
it seems that the quartessence Chaplygin gas models and LQG, can not be compatible with
each other. }. Secondly, by admitting
a down to up phenomenological view, our analysis gives qualitatively
a hint on the possibility of searching the micro-level spacetime via
the control of the Planck scale characteristic parameter using the
current astronomical observational signatures.} \\

At the end, to emphasize on the importance of the latter as an incentive for proposing an upcoming project, we would like
to refer to \cite{PLB} in which via probing the effects of NC geometry using the latest CMB
observations, authors have presented some positive feedbacks.\\

\section{Acknowledgment}
The authors would like to thank an anonymous referee for insightful comments. This work has been supported financially by Research
Institute for Astronomy and Astrophysics of Maragha (RIAAM) under research project number 1/5750-**.

\end{document}